\documentclass[10pt,oneside]{article}
\usepackage{geometry}
\usepackage{graphicx}
\usepackage{graphics}
\usepackage{amsfonts}
\usepackage{amsmath}
\usepackage[english]{babel} 
\usepackage{geometry} 
\usepackage{hyperref}
\usepackage{epstopdf}

\title{
	Big data Fusion to Estimate Urban Fuel Consumption: A case study of Riyadh
}

\author{Adham Kalila$^{1}$, 
Zeyad Awwad$^{2}$,
Riccardo Di Clemente$^{1}$,
\& Marta C. Gonz\'alez$^{1}$\thanks{\mbox{Corresponding author: \emph{E-mail~address}:~martag@mit.edu}}\\
\footnotesize
$^{1}$\textit{Department of Civil \& Environmental Engineering, MIT 77 Mass. Ave., Cambridge, MA 02139}\\
\footnotesize
$^{2}$\textit{Center for Complex Engineering Systems, King Abdulaziz City for Science and Technology, Riyadh 11442, Saudi Arabia}\\
}

\begin{document}

\date{}

\maketitle

\begin{abstract}
Falling oil revenues and rapid urbanization are putting a strain on the budgets of oil producing nations which often subsidize domestic fuel consumption. A direct way to decrease the impact of subsidies is to reduce fuel consumption by reducing congestion and car trips. While fuel consumption models have started to incorporate data sources from ubiquitous sensing devices, the opportunity  is to develop comprehensive models at urban scale leveraging sources such as Global Positioning System (GPS) data and Call Detail Records. We combine these big data sets in a novel method to model fuel consumption within a city and estimate how it may change due to different scenarios. To do so we calibrate a fuel consumption model for use on any car fleet fuel economy distribution and apply it in Riyadh, Saudi Arabia. The model proposed, based on speed profiles, is then used to test the effects on fuel consumption of reducing flow, both randomly and by targeting the most fuel inefficient trips in the city. The estimates considerably improve baseline methods based on average speeds, showing the benefits of the information added by the GPS data fusion. The presented method can be adapted to also measure emissions. The results constitute a clear application of data analysis tools to help decision makers compare policies aimed at achieving economic and environmental goals. 
\end{abstract}

\section{Introduction}

In many oil producing countries with substantial fuel subsidies, a fall in oil revenue and increasing domestic consumption has put increasing strain on government budgets \cite{kerr_2015}. Countries in the Gulf Cooperation Council, including Saudi Arabia and the UAE, have launched programs to reduce government expenditure on energy subsidies \cite{saudi_OBG_2016}. In Saudi Arabia, energy subsidies are estimated at 9.3\% of GDP, with 1.4\% for petroleum subsidies alone. Decreasing energy subsides can be achieved in a number of ways but congestion relief offers a simple and direct path to lower fuel consumption. As the burden of fuel subsidies continues to grow, it has become increasingly important for these countries to find simple and accurate methods to quantify the effects of policies on congestion relief and fuel consumption in cities. Recent technological advances in collecting and analyzing big data offer a potential method to measure how policy changes impact fuel consumption. With the advent of ubiquitous sensing devices, Transport Network Companies (TNCs), and new methods of estimating flow, we propose a method to answer such questions that can be further applied anywhere in the world and extended to model emissions and air pollution. 

Call Detail Records (CDRs) produced passively by mobile phones represent a cutting edge method to estimate travel demand. Most traffic studies currently use local and national household travel surveys to estimate the rate of trip production between different zones of the city but such surveys are expensive to conduct and only cover a small sample of the population. Leveraging on previous work \cite{ccolak2015analyzing, toole2015path, chodrow2016demand}, CDRs can provide simple and effective methods for estimating Origin-Destination (OD) flows using location data collected from millions of individual mobile phone users. 

In addition to CDRs, the Global Positioning System (GPS) function on smartphones offers a precise method to spatially and temporally track traffic conditions across a city. Since smartphone market penetration is almost complete in the TNC industry \cite{do2011smartphone}, GPS tracking has been successfully used to estimate air pollution \cite{de2013improving}, instantaneous fuel consumption \cite{lei2010microscopic,pelkmans2004development,song2009aggregate,VilaÃ§a2015,ericsson2006optimizing}, and traffic conditions \cite{work2008impacts,work2009lagrangian,donovan2015using,herrera2010evaluation,mazare2012trade}. Most studies that use GPS data to estimate fuel consumption have focused on individual user fuel consumption for route optimization\cite{song2009aggregate,ericsson2006optimizing}. In this paper, we employ the method described in \cite{toole2015path} to model trip demand, then fuse this result with location data from GPS data to calculate fuel consumption during peak and off-peak hours in Riyadh with higher accuracy than models that use average speed.

Fuel consumption and emissions models have been extensively developed in the literature \cite{bowyer1985guide,joumard1995hot,jimenez1998understanding,cappiello2002statistical,rakha2004development,ribeiro,zhou2016review,louhghalam2017carbon}. They are generally split between models that estimate the fuel consumption by balancing the engine's carbon intake and combustion and those that attempt to use mode-specific variables, such as speed and acceleration, to fit a model that estimates fuel consumption \cite{zhou2016review}. Of those models that use instantaneous mode-specific variables, some estimate air pollution and emissions\cite{de2013improving,joumard1995hot,jimenez1998understanding,rakha2004development}, fuel consumption \cite{song2009aggregate,VilaÃ§a2015,ericsson2006optimizing,ribeiro} or both \cite{lei2010microscopic,pelkmans2004development,cappiello2002statistical}. Most previous attempts at estimating instantaneous fuel consumption and emissions do not incorporate GPS data but rely instead on  On-Board Diagnostics devices (OBD-II) that measure fuel consumption and emissions \cite{lei2010microscopic,pelkmans2004development,song2009aggregate,cappiello2002statistical}. The models that have attempted to use GPS data to estimate fuel consumption do so without consideration to the different fuel efficiencies found in today's cars \cite{VilaÃ§a2015,ericsson2006optimizing,ribeiro} and do not account for the total demand.

With the aim of filling this gap, the contributions of this work are threefold: First, we calibrated a previously developed fuel consumption model (StreetSmart) and applied it to varying fuel efficiencies in car fleets; Second, the model is used along with travel demand estimated by CDR-based traffic assignment to approximate fuel consumption rates in Riyadh, Saudi Arabia; Third, we examined the effects of the proposed method comparing the reduction on fuel consumption in two scenarios: random trip reduction and reduction targeted at the least fuel efficient trips. We find that our proposed models calibrated with speed and acceleration profiles considerably differ in their fuel consumption estimates at the city scale when verified against estimations from a baseline model with average speeds. 

The paper is structured as follows. In section 2.1 we discuss the methodology used to calibrate the fuel consumption model (StreetSmart). In section 2.2, we describe the cleaning of the GPS data and the extraction of speed and acceleration profiles on each street in specific time windows in a typical week to represent different snapshots of traffic throughout the road network of Riyadh. In section 2.3 we describe the application of the model combined with flow data to visualize the fuel consumption across different time periods. Moreover, we present the results of the targeted and random flow reductions and their effects on fuel consumption via the presented model vs. baseline estimates. Finally,  in Section 3 we discuss the results and the conclusions derived from the study.

\section{2 Methodology}
\subsection{2.1 StreetSmart Model Calibration}
 
In this work, we use the StreetSmart model to estimate the fuel consumption of personal vehicles \cite{oehlerking2011streetsmart}. It was comprehensively developed by measuring the energy required by vehicles for various movement conditions. It estimates the fuel consumption with data from GPS coordinates from smartphones and ground truth fuel consumption data from OBD-II devices . Using the details of a trip's speed profile, the model successfully predicts fuel consumption with over 96\% accuracy \cite{oehlerking2011streetsmart}, a substantial improvement over models that only consider constant average speeds. Average speed estimations do not account for the stop and go effect of traffic, which is a significant factor leading to an increase in fuel consumption. In other words, average speed simplification results in lower, or more optimistic, fuel consumption estimates since drag is lower at low speeds. 

After testing different variables for their use in predicting fuel consumption, the model employs a combination of four variables to predict fuel consumption as shown in equation \ref{eq:1}. The first term accounts for energy wasted while the car is idling with the engine turned on; the second accounts for energy used with time spent moving; The third accounts for energy used due to acceleration and deceleration over a distance; the fourth accounts for energy used with distance traveled. Each term is multiplied by a specific energy index, $k_i$, which depends on the vehicle efficiency, such that: 
\begin{equation} \label{eq:1}
FC = k_1T_{idle} + k_2T_{move} + k_3\int |a|dx + k_4L,
\end{equation}
where FC is fuel consumption in US $gal$ and $k_1, k_2, k_3,$ and $k_4$ are the energy indices calibrated with data separately for each bin of vehicle efficiency. $T_{idle}$ and $T_{move}$ are time spent idling and moving respectively in $sec$, $a$ is acceleration in $m/s^2$, and $L$ is the distance driven in $km$. 

\begin{figure}[h!]
\centering
\includegraphics[width=.8
\linewidth]{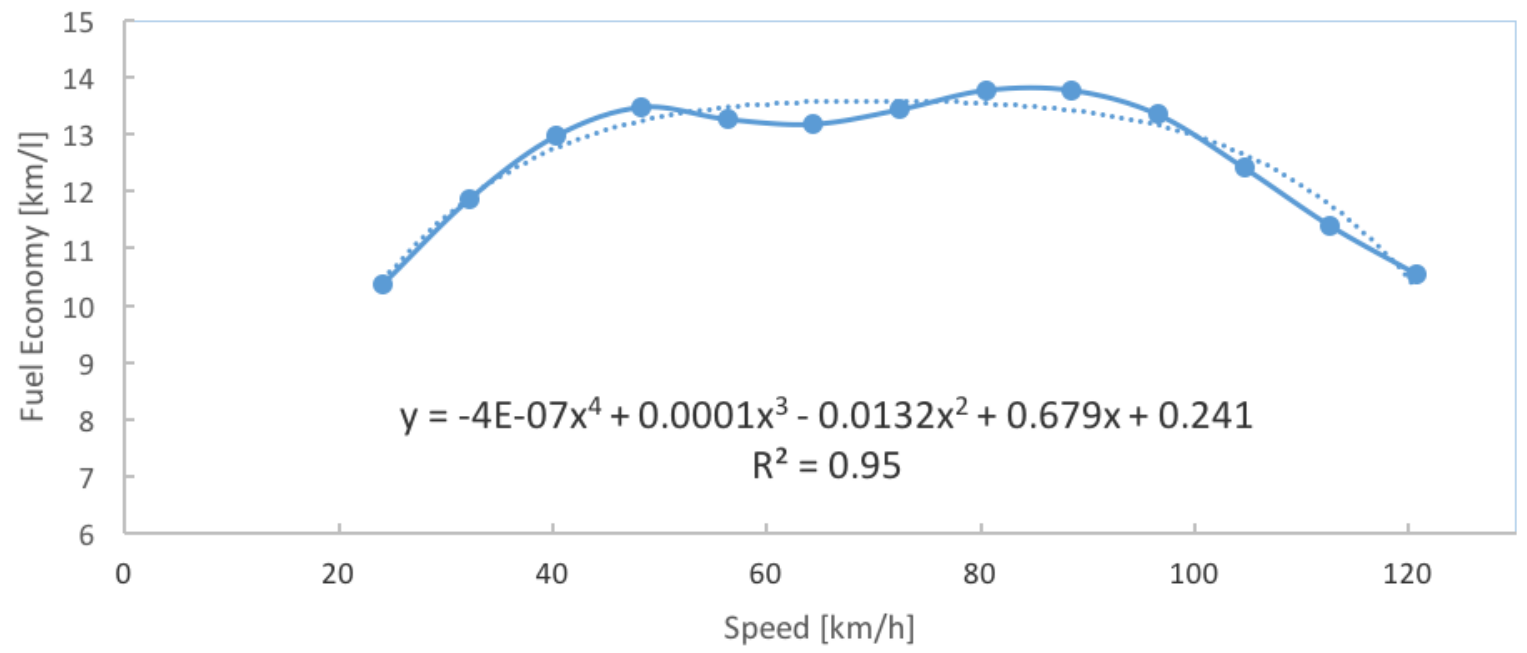}
\caption{A fitted model of fuel economy vs. speed from DOE as a baseline comparison to the StreetSmart model. \cite{doe-book} }
\label{fig:0}
\end{figure}

Following \cite{oehlerking2011streetsmart}, we further tested the StreetSmart model's indices by regression of the idle fuel consumption and moving fuel consumption separately using data from an experiment conducted at the University of Illinois by \cite{wu2017measuring}.  To verify the benefit of using the model, its results were compared to a baseline estimation using average speed and the United States' Department of Energy's (DOE) graph of fuel economy variations by speed as shown in Figure \ref{fig:0}. The StreetSmart model achieved inaccuracies of about 4\% while the baseline method had inaccuracies of up to 29\%. The details of the comparison are summarized in Table \ref{tableSS}.

\begin{figure}[h!]
\centering
\includegraphics[width=.99\linewidth]{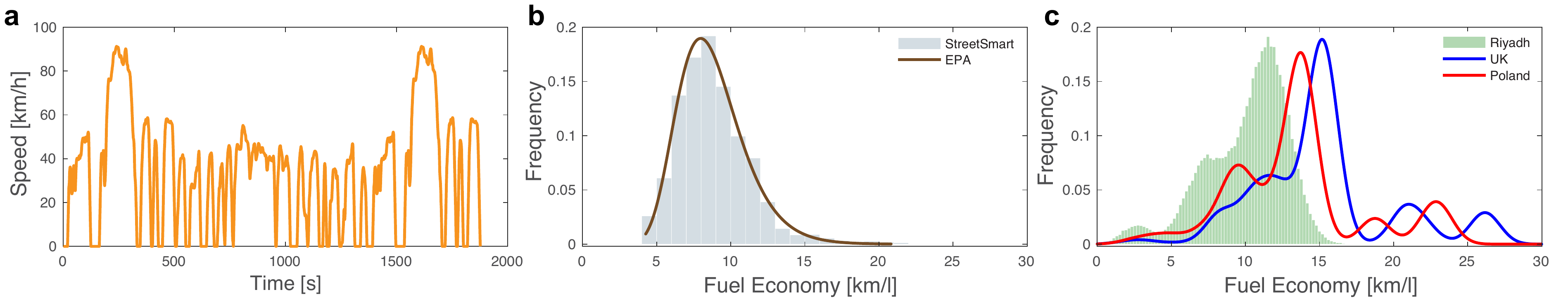}
\caption{(a) FTP-75 EPA's standard speed profile used for calculating the reported inner-city fuel economies of cars. (b) The distribution of fuel economies recreated by the StreetSmart model shows the same distribution as that of the reported fuel economies. (c) The distribution of fuel economies based on Riyadh's fleet of cars compared to those of Poland and the UK shows that the distributions are similar but shifted from one another. The car fleet of Riyadh is less fuel efficient than that of Poland which is less than that of the UK}
\label{fig:1}
\end{figure}

To apply the StreetSmart model on a known distribution of car fuel economies, the first step is to relate the energy indices specific to each car type in order to predict fuel consumption. We conducted a sensitivity analysis to assess the influence of each term on the estimates and found that the fuel consumption estimates were most affected by the second and fourth terms. This indicates that the variables with the highest influence over fuel consumption were the time moving and the distance traveled. This suggests that the first and third terms, representing the influence of speed profiles, have little impact on the overall estimate of fuel consumption. However, results shown in section 2.3 show that this is not the case. To calibrate the model and get energy index values for use on the scale of a city, we used the fuel economies reported by the Environmental Protection Agency's (EPA) 2016 report to arrive at ranges of each index for different cars, categorized by their fuel economy \cite{fuel_guide_2016}. The EPA uses a standard speed profile to test for a car's urban fuel economy, the FTP-75, shown in Figure~\ref{fig:1}a~\cite{epa_2017}. We used the mode specific variables from the FTP-75 speed profile with the reported fuel economy to calibrate the energy index ranges for each bin shown in Table \ref{table1:kranges}. Using these ranges and the distribution of fuel economies found in the EPA report, we successfully recreated the distribution of fuel economies (shown in in Figure \ref{fig:1}b) with the SmartStreet model . 
\begin{table}[]
\centering
\caption{A Comparison of Fuel Consumption Estimates from the StreetSmart Model  and the DOE Fuel Economy Fit on Data From the Experiment Conducted by \cite{wu2017measuring})}
\label{tableSS}
\begin{tabular}{lllll}
       &        &        &         &        \\
         Car No. (Illinois Test A) :  & 6      & 7      & 8       & 9      \\
\hline
OBD-II FC {[}US Gal{]}      & 0.0246 & 0.0220 & 0.0356  & 0.0211 \\
DOE Fitted Curve  {[}US Gal{]} & 0.0247 & 0.0253 & 0.0252  & 0.0245 \\
StreetSmart FC {[}US Gal{]} & 0.0243 & 0.0230 & 0.0371  & 0.0210 \\
\% diff. StreetSmart        & -1.2\% & 4.2\%  & 4.2\%   & -0.6\% \\
\% diff. DOE Fitted Curve   & 0.4\%  & 14.7\% & -29.3\% & 16.0\%
\end{tabular}
\end{table}

The energy index ranges for each bin are then used to recreate the fuel economy distribution of the fleet of cars in Riyadh using the car makes and models from motor vehicle crash statistics data provided by the city of Riyadh. Data on car crashes from January 2013 until October 2015 in the city of Riyadh were used as a proxy for Riyadh's fleet composition. The different energy indices, which produce different fuel consumption rates, are combined in proportion to Riyadh's fleet bin distribution. For validation, the fuel economy distribution of Riyadh's fleet was compared to two cities in Europe chosen on the basis of similar population size or Gross Domestic Product and the results show similar trends exist in the relative variety of fuel efficiencies in all cities. The data on fleet compositions of the two European countries were acquired from an in-depth study of the fleets of all European countries \cite{traccs}. As can be seen in Figure \ref{fig:1}c, the comparison in the distributions of fleet fuel economies between the three areas indicates that the car fleet of Riyadh is less energy efficient those of Poland and the UK.  The usage of country level fleet composition for Poland and the UK compared to city level for Riyadh represents a limitation in this comparison but the results adequately verify the credibility of the fleet composition used in this study. With the relative fuel economies of Riyadh's fleet and the calibrated StreetSmart model, we discuss next how we integrate speed profiles from GPS data to estimate fuel consumption at the urban scale.

\begin{table}[]
\centering
\caption{Results of the Calibration of the StreetSmart Model. Ranges of $k_i$ Parameters for Each Bin of Fuel Economy}
\label{table1:kranges}
\resizebox{\textwidth}{!}{\begin{tabular}{lllllll}
Bin             & FE Range {[}MPG{]} & FE Range {[}km/l{]} & $k_1$          & $k_2$          & $k_3$      & $k_4$\\
\hline
1               & {[}10 - 12)        & {[}4.25 - 5.10)     & 37.0        & 30 - 34     & 1 - 4.8 & 2000 - 2300 \\
2               & {[}12 - 14)        & {[}5.10 - 5.95)     & 34.6        & 23 - 32     & 1 - 4.8 & 1300 - 2300 \\
3               & {[}14 - 16)        & {[}5.95 - 6.80)     & 31.9        & 21 - 26     & 1 - 4.8 & 1100 - 1900 \\
4               & {[}16 - 18)        & {[}6.80 - 7.65)     & 29.5        & 17.5 - 24   & 1 - 4.8 & 1000 - 1600 \\
5               & {[}18 - 20)        & {[}7.65 - 8.50)     & 26.9        & 15 - 22     & 1 - 4.8 & 1000 - 1250 \\
6               & {[}20 - 22)        & {[}8.50 - 9.35)     & 24.3        & 13 - 18     & 1 - 4.8 & 980 - 1250  \\
7               & {[}22 - 24)        & {[}9.35 - 10.20)    & 21.7        & 12 - 16     & 1 - 4.8 & 850 - 1200  \\
8               & {[}24 - 26)        & {[}10.20 - 11.05)   & 19.0        & 12 - 15     & 1 - 4.8 & 750 - 1050  \\
9               & {[}26 - 28)        & {[}11.05 - 11.90)   & 16.3        & 11 - 14     & 1 - 4.8 & 780 - 900   \\
10              & {[}28 - 30)        & {[}11.90 - 12.75)   & 14.0        & 10.5 - 12.5 & 1 - 4.8 & 710 - 900   \\
11              & \textgreater 30    & \textgreater 12.75  & 5.0         & 5 - 14.5    & 1 - 4.8 & 500 - 1000  \\
12 (Bus)        & 6.3                & 2.68                & 30.0 - 37.0 & 30 - 75     & 1 - 4.8 & 2000 - 8000 \\
13 (Truck)      & 17.27              & 7.34                & 29.0 - 30.0 & 12 - 27     & 1 - 4.8 & 500 - 2200  \\
14 (Motorcycle) & 43.5               & 18.49               & 5.0         & 6 - 10      & 1 - 4.8 & 500 - 600  
\end{tabular}}
\end{table}

\subsection{2.2 From GPS Data to Speed Profiles}
 
\begin{figure}[h!]
\centering
\includegraphics[width=.9\linewidth]{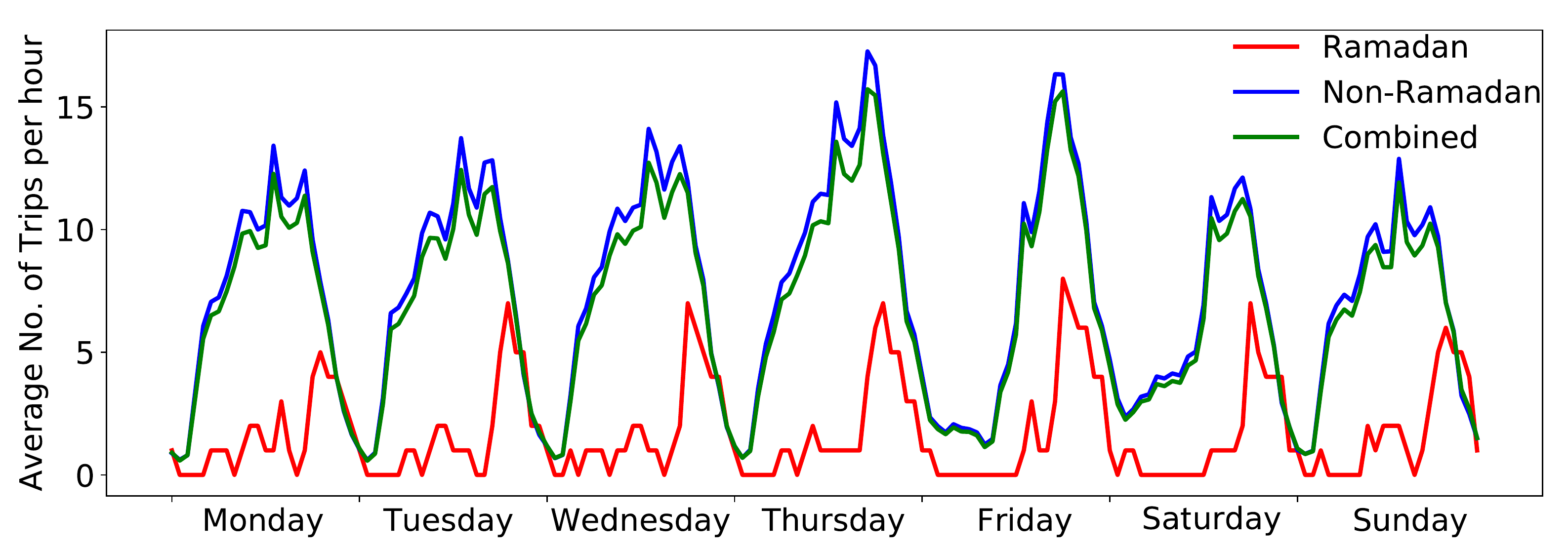}
\caption{Average hourly taxi trip production rates in Riyadh in Ramadan, non-Ramadan, and combined.}
\label{fig:2}
\end{figure}

We extracted speed profiles from a large dataset of GPS tracking points of taxi trips from a local Saudi Arabian TNC company over the period of May 2015 until December 2016. Speed profiles are the result of both driving styles and traffic conditions. Driving styles of taxis may be different from those of personal vehicles, affecting the results in a small extent, but we assume speed profiles are mainly a reflection of traffic conditions which would affect taxis and personal vehicle trips similarly. The dataset included trip duration and length, pick up and drop off times, and a chronologically ordered list of GPS coordinates. To ensure that the traffic is representative of year-round conditions we compared the rate of trip production during Ramadan of 2015 and 2016 with non-Ramadan trip production rates. We found that Ramadan trip production rates are much fewer so their impact on the average traffic speed profile for a specific street is negligible. For this reason, we kept the Ramadan trips in the analysis to benefit from the higher amount of data on the street level. A graph of the average number of trips per hour during Ramadan, non-Ramadan, and combined can be seen in Figure~\ref{fig:2}. 

Before use, the data was filtered to remove trips that were outside Riyadh and the GPS routes were cleaned and modified to correct for measurement errors. We detected errors in the GPS points and fixed them using the following algorithm. Since the GPS coordinates were given as an ordered list without a timestamp, and are known to be collected at regular intervals, the interval or frequency of recording was calculated as the total duration of the trip divided by the number of points recorded. Errors were detected as spikes in speed that are greater than 160 km/h since the taxi's fleet and the road conditions typically preclude speeds above that. Two different causes of error were detected and repaired using different methods. In the first case, errors were caused by missing points due to a lack of network signal. This would result in a spike in speed for one segment but not the following segment, which would revert to realistic speeds and GPS points. The number of skipped points was estimated from the average of the speed before and after the single speed spike. Second, for errors caused by a GPS point that is in an obviously anomalous location, the speed spike occurs in two simultaneous segments, one to jump to the wrong location and another to return to the realistic location. This error was fixed by removing the erroneous point. This simple method was able to adequately correct the GPS coordinates.

We also cleaned the taxi data using an algorithm developed by \cite{jiang2013review} to detect long periods of immobility, or stays, that can be interpreted as parking in our dataset. This may occur if a client asks to keep the meter running while they finish an errand. The reason for the splitting is not to affect the recorded speeds on the streets where the taxi is effectively parked. We chose the minimum thresholds for idling time and distance by inspecting the distribution of stay durations throughout the week. A minimum value of $2,200 s$ (around 37 $min$) was chosen to allow for the majority of traffic stays during peak and off-peak hours. A maximum distance for a trip to be split around a stay was 10 m. This number reflects the relatively high precision of the GPS points. In other words, a trip that remained within a 10 $m$ radius for longer than 37 $min$ was split into two trips, removing the 37 $min$ or longer section of immobility.

After filtering and cleaning GPS routes for the city of Riyadh, we obtained nearly $43,000$ trips that were analyzed by a custom mapping algorithm to assign full GPS routes to edges in the road network. For more accurate mapping, longer edges in the road network were split into shorter segments to ensure that nodes are no more than 10 $m$ apart. The mapping algorithm was implemented using the following procedure:

\
\begin{enumerate}
\item For each point $i$ in the GPS trajectory, we identify the set of nodes ($N_i$) in the road network that fall within a 25 m radius.
\item We constructed a path network $G$ consisting of the nodes $N_i$.
\item For each point $i$ in the GPS trajectory, we used the Dijkstra algorithm to find the fastest route from every node in $N_i$ to every node in $N_{i+1}$. For each route, we added a representative edge to $G$ with the route's total travel time as the weight. 
\item For each edge, we added a time penalty based on the distance of the target node to the original GPS coordinate at a rate of 1 second per additional meter past the closest node. 
\item Any gaps in $G$ were identified to determine contiguous sequences of paths that represent segments of potential routes.
\item For each contiguous sequence, we identified the fastest path in $G$ as the most likely route taken by the vehicle.
\end{enumerate}

\begin{figure}[h!]
\centering
\includegraphics[width=1.0\linewidth]{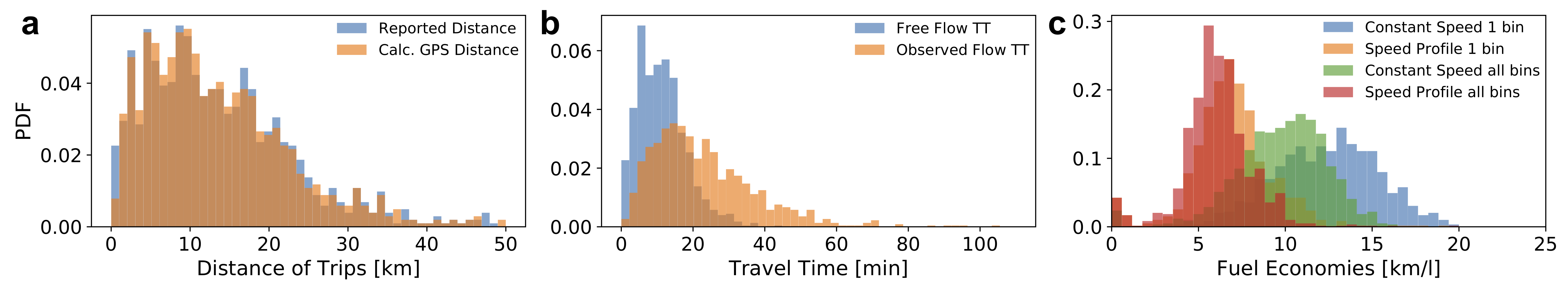}
\caption{Data Verification figures using trips in the morning peak time period of weekdays from 8 - 9 AM. (a) Histogram of Reported and calculated Trip Distances. (b) Histogram of free flow travel time and Observed travel time in matched trips. (c) Histogram of Fuel economies using constant speed, speed profiles, and 1 bin and all bins.}
\label{fig:3}
\end{figure}

For verification purposes, trip distances, free flow and observed travel times as well as fuel economy estimates are plotted in Figure~\ref{fig:3}. Fuel economy is defined as the distance traveled per liter of fuel consumption. In Figure~\ref{fig:3}a, we verified that the reported distances were generally consistent with the sum of the distances calculated between every two consecutive GPS points in each trip using the Haversine formula. In Figure~\ref{fig:3}b the free flow times, computed as the sum of the free flow times of every matched segment in each trip is compared to the observed flow times as reported in the taxi data.  The comparison only considered the observed total times from trips that were successfully matched with streets. The figure shows consistent results with observed travel times during the morning peak hour being slower than the free flow times of each trip. As a baseline comparison to using speed profiles from taxi data, we calculate fuel economy with the StreetSmart model assuming a constant speed and one fuel economy bin based on the Hyundai Elantra, the most common car according to the accident data ( Figure~\ref{fig:3}c). We plotted the results if speed profiles are used with only one fuel economy bin and if all bins are used in proportion to the fleet of Riyadh from the crash statistics data. It shows a very close distribution to the simpler assumption that all cars are the most common car fuel economy, and in high contrast to the result given by not using the speed profiles. Incorporating speed profiles in the model results in higher fuel consumption, or much lower fuel economy, which is more accurate and important for policy projections. This constitutes the core benefits of integrating the more accurate fuel consumption model.

\subsection{2.3 Fuel Consumption Results}
 
With the StreetSmart model calibrated and speed profiles extracted from GPS data, we can now estimate fuel consumption for three typical time periods representing distinct traffic conditions. The time periods are morning peak (8 - 9 AM), midday off-peak (12 - 13 PM), and evening peak (17 - 18 PM) during weekdays. For every edge in the road network, we calculated fuel consumption per car based on each speed profile matched to that edge. The estimates were computed for each bin of fuel economy in the Streetsmart model.  For each edge, the average fuel consumption rate per car was multiplied by the flow of cars per hour as computed by a version of the Iterative Traffic Assignment (ITA) algorithm to get a fuel consumption rate per hour, with car volumes included. 

The flow was calculated from OD matrices previously derived from CDR data in Riyadh by \cite{toole2015path}, in cars/hour over morning, midday, and evening time periods. 
Following \cite{toole2015path}, a factor of 1.5 was applied to the average morning flow of the morning and evening time periods to determine peak hour demand. The congested time estimates correlated successfully with the ones acquired from Google Maps (see comparisons in ~\cite{chodrow2016demand}).

\subsubsection{Fuel Consumption Rate }  
As presented in ~\cite{chodrow2016demand}, the ITA algorithm assigns trips in a series of four increments. It incrementally assigns 40\%, 30\%, 20\% and 10\% of the OD flows to the fastest routes in the road network. After each iteration, the congested time of each edge is updated so that the effects of congestion can be factored into the assignment. The resulting flow volume is multiplied by the fuel consumption rate per car to arrive at hourly fuel consumption as shown in Equation \ref{eq:3} which is normalized by edge length.
\begin{equation} \label{eq:3}
FC = \frac{flow_e[\frac{car}{hr}]\times fcr_e[\frac{liter}{car}]}{L_e[m]},
\end{equation}
where FC is the rate of fuel consumption in $[\frac{liter}{m.hr}]$, $flow_e$ is the flow on edge $e$ as estimated by the assignment algorithm, $fcr_e$ is the rate of fuel consumption per car on edge e as estimated by our application of the StreetSmart model, and $L_e$ is the length of the edge in meters. 

The results of the model are shown in the choropleth maps in Figure~\ref{fig:4} below. For the fuel consumption rate per street, we used a weighted average of fuel consumption by bin proportional to Riyadh's fleet.
The maps show that the most fuel consuming streets are the grid highways of the city. As expected, the flow of cars and total fuel consumption per meter of road is found to be higher in the peak periods of the morning and evening than the midday off-peak. The evening peak shows slightly higher fuel consumption than the morning peak, indicated by more red streets. We used quantile breaks on the fuel consumption rate values of all time periods combined to display the streets that are the most fuel consuming. A high fuel consumption rate per meter of road is due to high fuel consumption rate per car and high car flow values. We observe that King Fahd Road is by far the most fuel consuming road in the city, especially in the area bounded by the old city center from the south and the Northern Ring road from the north.
\begin{figure}[h!]
\centering
\includegraphics[width=1.0\linewidth]{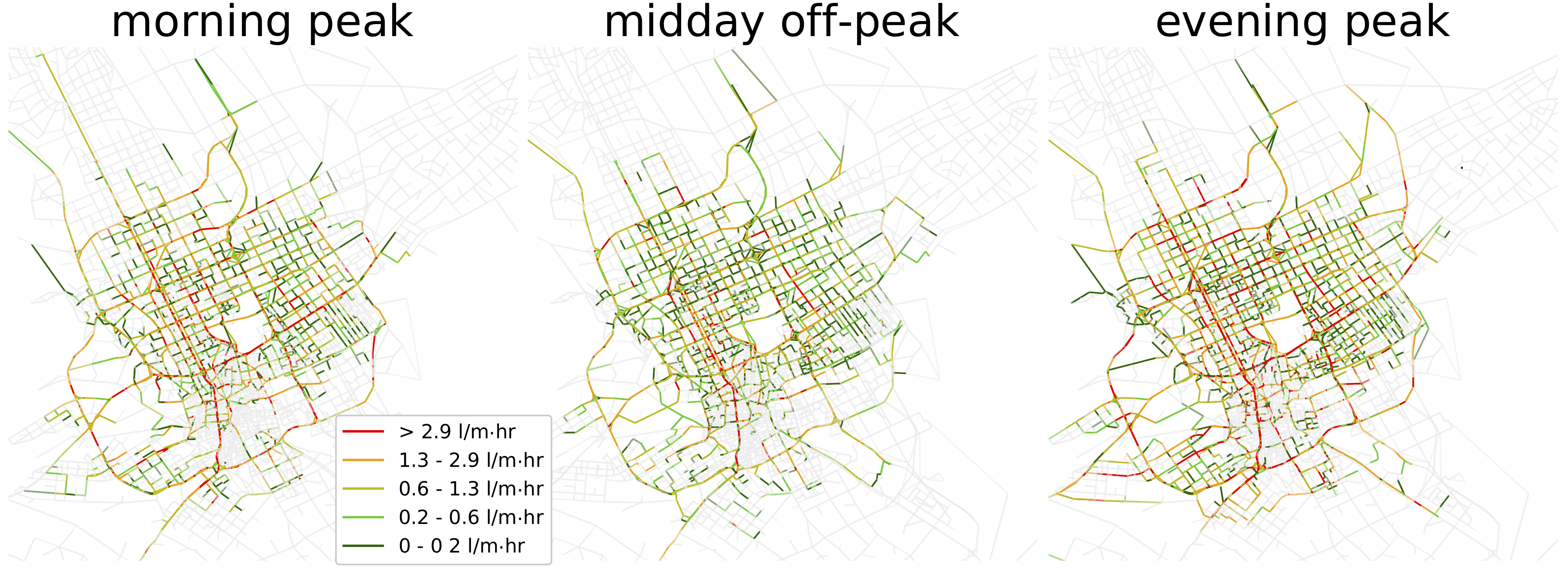}
\caption{Choropleth Maps of fuel consumption rates [Liter/meter.hour] by the StreetSmart model on streets matched with GPS data for typical time periods morning peak (8 - 9 AM) weekdays, midday off-peak (12 - 13 PM) weekdays, evening peak (17 - 18 PM) weekdays.}
\label{fig:4}
\end{figure}

The taxi GPS data necessary to calculate a fuel consumption on each street did not cover the entire network of streets. Similarly, the traffic assignment of ODs did not use all of the city's streets and the majority of roads used were also matched with at least one speed profile from the taxi GPS data. Specifically, 56\%, 57\%, and 63\% of streets were covered for the morning,midday, and evening time periods respectively. 
\subsubsection{Relevance of integrating GPS data}  
\begin{figure}[h!]
\centering
\includegraphics[width=1.0\linewidth]{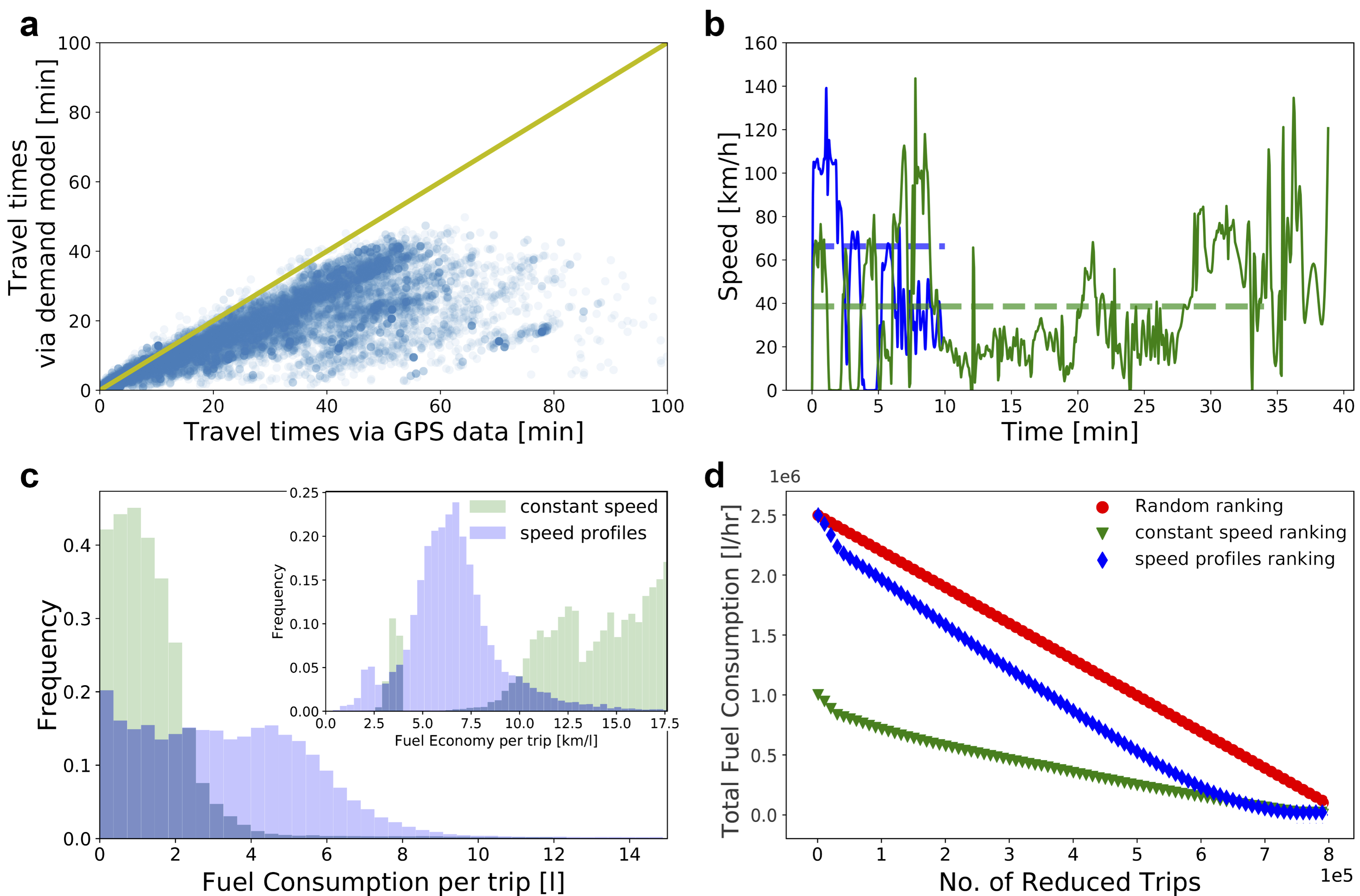}
\caption{Fuel consumption estimates at urban scale.
(a) Comparison of the travel time of the routes in the constant speed model via travel demand vs. the input used in our method using GPS data (b) Sample speed profiles in two routes used to estimate fuel consumption overlaid with the constant speed used for comparison (c) Estimates of fuel consumption and fuel economy in the morning peak via our method (speed profiles) and the base line method (constant speed)  (d)  Random and targeted fuel savings vs. number of reduced trips.}
\label{fig:5}
\end{figure}

For the morning time period, a simulation of all OD trips in the city is made and the path, defined as the sequence of edges taken from origin node to destination node, is defined using the shortest path algorithm as in the ITA algorithm. In the simulation, OD pairs with flow values that are less than one car/hour are omitted to ensure each flow represents a discrete trip. We assign a car fuel economy bin at random to each trip in proportion to the probability of that bin in Riyadh's fleet. Using the StreetSmart model, we estimate fuel consumption and trip time as the sum of the their values on each edge in the trip's path. For verification, total trip times derived via the demand model which were previously verified against Google Maps estimates \cite{chodrow2016demand} are plotted against the trip times derived via the GPS data. As shown in Figure \ref{fig:5}a, travel times via GPS data are generally higher than their counterparts from the simulation but their correlation is satisfactory.

Two sample speed profiles used by the StreetSmart model are plotted with the constant speed assumed by the baseline comparison. As expected, not all trip times are equivalent to their speed profile counterparts and the constant speed is generally lower than the peaks of the speed profile. Extracted speed profiles do not always end at 0 $km/h$ speed since the originally matched trips on these edges may not be stopping at that edge. This overestimation of speed at the end of a trip represents a negligible increase in fuel consumption estimation.  The lower constant speed assumption used in the baseline comparison results in fuel consumption estimates that are consistently lower than those from using speed profiles from GPS data. Thus, the effect of the accelerations and deceleration on the StreetSmart model are observed to be significant and not negligible which shows the benefit of using GPS data in the fuel consumption model. The distributions of fuel consumption per trip using speed profiles and constant speed are shown in Figure \ref{fig:5}c along with the fuel economies. It is clear that the acceleration and detailed speed profiles are not negligible in the estimates. These results bring to the urban scale the results of Figure \ref{fig:3}c. 

The effect of reducing flows on overall fuel consumption is shown in figure ~\ref{fig:5}d. Trips were removed from the simulation described above via three rankings and the resulting fuel saving potential is shown, normalized by the total fuel consumption of each method. Random trip reduction results in a perfectly linear fuel saving effect. In contrast, we see the best case scenario, where trip reduction of the worst fuel consuming trips per meter are ranked. As expected, targeted reduction results in higher fuel savings with the same number of reduced trips. More importantly, when comparing the speed profile estimates (method proposed here) vs. constant speed fuel estimates (baseline) differences emerge.  Constant speed targeting shows a higher return on fuel savings because the variance of the distribution of fuel consumption estimates using constant speed is less than that of the speed profiles. In other words, the fuel economy distribution of the constant speed estimates shows higher proportions at the high end of the distribution than the high end of the speed profile distribution shown in Figure \ref{fig:5}c. This would explain the higher fuel savings observed when the trips are ranked by constant speed fuel consumption per meter. However, since the constant speed estimates are not as accurate as those derived using speed profiles, the fuel savings are spurious. 

The real gain in using speed profiles to estimate fuel consumption over the baseline constant speed assumption is in the city-wide total fuel consumption per hour estimate. The constant speed assumption underestimates the city-wide fuel consumption for the morning hour by 60\% compared to using the speed profiles.  

The relative gain in fuel saving potential of targeting the highest fuel consuming trips over random trip reduction is approximately 10\% if 14.5\% of trips are reduced. In other words, if 14.5\% of trips are reduced, the policy targeting the highest fuel consuming trips per $m$ would save 10\% more fuel for every morning peak hour. When 14.5\% of targeted trips are reduced, 25\% of fuel consumption for the morning peak hour is saved. These results are encouraging but the ratio is not as high as the effect of similar targeting policies on household energy consumption \cite{qomi2016data}, where the increase over random is  51\%. These differences can be explained by the more normal distribution of fuel economy of the trips which renders the effects of targeting to be much less dramatic than the more broad distribution found in house energy consumption. 

\section{Discussion and Conclusion}
We present a data fusion method to estimate fuel consumption at the urban scale. We leverage a travel demand model that uses mobile phone data and integrate speed profiles from taxi GPS data that covered most of the street network of Riyadh. To identify fleet distribution, we used car crash statistics data as a proxy with the assumption that the distribution of car makes and models is representative of Riyadh's car fleet distribution. The method developed here and the calibration of the StreetSmart model for fuel consumption can easily be extended to any other region. It is significantly faster than if speed profiles were simulated by software which would require a large amount of time and computing resources \cite{cetin2003large}. 

The fuel consumption model was verified to produce better results than than the DOE fuel economy by speed graph and tested on real OBD-II fuel consumption measurements before being calibrated to be used on any car fuel economy. The resulting calibration results are presented in Table \ref{table1:kranges}. We used the calibrated StreetSmart model with speed profiles from GPS data to produce estimates which were compared with the baseline without GPS data and assuming constant speed. The results show that  the differences of using speed profiles is significant, justifying the introduction of the more elaborate model in policy estimates. 

As a proof of concept, we applied the calibrated StreetSmart model to test a policy of flow reduction, both random and targeted to the least fuel efficient trips and simulated the effect on the rate of fuel consumption in the city. We showed that the difference in fuel consumption reduction between targeted and random schemes was around 10\% more fuel savings for 14.5\% trip reduction. Interestingly, 25\% of city-wide fuel savings potential can be achieved by removing only 14.5\% of trips ranked by the worst fuel consuming trips per meter . 

While this project has demonstrated the potential of data-driven models to estimate the effects of policies on fuel consumption, it can benefit from further study to understand the impacts of several simplifications and assumptions. There are three main areas which require further research:
\begin{enumerate}
\item To correctly assess the benefit of our straightforward approximation over the computationally costly alternative, speed profiles of each car across every origin-destination trip should be simulated and the results compared.
\item The GPS data used in our model covered most but not all of the street network. A more accurate fuel consumption calculation can be achieved with more data, covering a higher proportion of flow over a longer time period. 
\item The trip assignment method used is an efficient and reasonably effective method to estimate overall congestion based on a simplified method of route choice. However, the baseline estimates would benefit from trip assignment derived from a dynamic traffic assignment model.
\end{enumerate}

We hope the method used to obtain the results of the fuel reduction strategies can be implemented for other fuel or emissions reduction strategies, assuming the conditions remain relatively unchanged. For example, the composition of the Riyadh fleet remains similar and that streets remain unchanged. For future implementations, the composition of the fleet can easily be adjusted since the method uses bins to account for all car types. However, a significant change in the road network layout or the traffic conditions would require new datasets (Road network and GPS routes for speed profile extraction) to recompute modular road-by-road fuel consumption estimates for each bin of car type.

Current fuel consumption models are adapting to the wealth of data available from ubiquitous sensing devices such as smart-phones. Our project aims to show the relative gain in accuracy of incorporating both fuel economy distribution considerations and speed profiles derived, in our case, from GPS devices used by a local taxi company. The method developed in this paper can be adapted to also measure emissions. It can be further augmented to analyze the economic and environmental impacts of policies targeting specific trips by conducting a network analysis to identify the affected trips. The tools developed in this work have the potential to assess the consequences of a variety of policies under different circumstances and in any region of the world.

\section{Acknowledgements}
 
The research was supported by grants from the MIT Energy Solutions Initiative, and from Center for Complex Engineering Systems at the King Abdulaziz City for Science and Technology.

\section{GitHub Repository}
 
The methods and code used to perform the analysis as well as the figures are available for public access at \url{www.github.com/adhamkalila/riyadhfuelconsump}. We encourage the scientific community to reproduce the analysis and notify us of any issues on the GitHub repository directly or by email.

\bibliographystyle{ieeetr}


\end{document}